\documentstyle[12pt,aaspp4]{article}
\def\Ginga{\hbox{\it Ginga}}
\tighten

\lefthead{Band et al.}
\righthead{GRB Line Detectability}

\begin{document}

\title{BATSE GAMMA-RAY BURST LINE SEARCH: \\ 
V. PROBABILITY OF DETECTING A LINE IN A BURST}
\author{D. L. Band, L. A. Ford\altaffilmark{1}, J. L. Matteson}
\affil{CASS 0424, University of California, San Diego, La Jolla, CA 
92093} 
\altaffiltext{1}{Present address:  Department of Physics and Astronomy,
University of Wisconsin, Eau Claire, P. O. Box 4004, Eau Claire, WI 
54702} 
\author{M. S. Briggs, W. S. Paciesas, G. N. Pendleton, R. D. Preece}
\affil{University of Alabama at Huntsville, Huntsville, AL 35899}
\centerline{\it Received 1996 October 18; accepted 1997 March 20} 
\centerline{To appear in {\it The Astrophysical Journal}}
\begin{abstract}
The physical importance of the apparent discrepancy between the
detections by pre-BATSE missions of absorption lines in gamma-ray
burst spectra and the absence of a BATSE line detection necessitates a
statistical analysis of this discrepancy.  This analysis requires a
calculation of the probability that a line, if present, will be
detected in a given burst.  However, the connection between the
detectability of a line in a spectrum and in a burst requires a model
for the occurrence of a line within a burst. We have developed the
necessary weighting for the line detection probability for each
spectrum spanning the burst. The resulting calculations require a
description of each spectrum in the BATSE database. With these tools
we identify the bursts in which lines are most likely to be detected. 
Also, by assuming a small frequency with which lines occur, we
calculate the approximate number of BATSE bursts in which lines of
various types could be detected.  Lines similar to the \Ginga\
detections can be detected in relatively few BATSE bursts; for
example, in only $\sim 20$ bursts are lines similar to the GB~880205
pair of lines detectable.  \Ginga\ reported lines at $\sim20$ and
$\sim40$~keV whereas the low energy cutoff of the BATSE spectra is
typically above 20~keV; hence BATSE's sensitivity to lines is less
than that of \Ginga\ below 40~keV, and greater above. Therefore the
probability that the GB~880205 lines would be detected in a \Ginga\
burst rather than a BATSE burst is $\sim0.2$.  Finally, we adopted a
more appropriate test of the significance of a line feature. 
\end{abstract}
\clearpage
\section{INTRODUCTION}
The continued absence of a line detection in the gamma-ray burst
spectra accumulated by the Burst and Transient Source Detector (BATSE)
on the {\it Compton Gamma-Ray Observatory (CGRO)} (Palmer et al. 1994;
henceforth Paper~I) has led us to continue not only the search for
lines in the BATSE data (Briggs et al. 1996), but also our study of
the detectability of lines by the BATSE detectors and the statistical
implications of the current results.  In particular, we are evaluating
the consistency between the BATSE observations and those of previous
missions, particularly those of \Ginga. These calculations assume the
BATSE detectors function properly and that our models of their
performance are accurate, assumptions which we test continuously
(Paciesas et al. 1997; Preece et al. 1997).  Here we fill a major gap
in our statistical methodology and implement it for the BATSE data. 

The description of our statistical methodology is clearest using
conditional probabilities and their associated notation.  Thus $p(a
\,|\, b)$ means the probability of proposition $a$ given proposition
$b$. A bar over a proposition denotes the negation of that
proposition.  Since we need to differentiate between quantities which
refer either to a burst as a whole or to a specific spectrum
accumulated over a portion of the burst, we use the convention that
roman indices specify spectra and greek indices identify bursts.  For
example, $l_\sigma$ represents the proposition that a line exists in
the $\sigma$th burst, while $l_i$ is the proposition that a line is
present in the $i$th spectrum; technically the burst within which the
spectrum was accumulated should also be indicated (e.g., $l_{\sigma
i}$), but the burst will be understood from the context.  As a
reminder that these probabilities rely on our understanding of the
detectors and gamma-ray bursts we include as one of the givens the
proposition $I$, which represents our model of the detector response,
our parameterization of the burst continuum, etc. Our calculations can
be seriously in error if our assumptions expressed by $I$ are
incorrect. For example, we use the ``GRB'' spectral function (Band et
al. 1993) to model the continuum, but this spectral shape is not based
on the source physics, and therefore must be incorrect at some level
of accuracy. 

Our analysis of the possible line content of a burst sample is based
on a hierarchy of probabilities.  Ultimately we want the probability
$p(D \,|\,HI)$ of obtaining the observed data (proposition $D$)
assuming hypothesis $H$ (Band et al. 1994, hereafter Paper~II).  Thus
$D$ might represent the statement that no lines have been detected in
the BATSE database, and $H$ might be the hypothesis that lines exist
and that we are modeling BATSE correctly.  The information which might
be represented by $I$ and $H$ overlaps; in general $H$ should include
the information which differs when hypotheses are compared. Also known
as the likelihood for $H$, $p(D \,|\,HI)$ can be used in measures of
the consistency between BATSE and previous detectors.  Our methodology
does not result in $p(D \,|\,HI)$ directly, but rather in $p(D
\,|\,fHI)$ where $f$ is the line frequency, the probability that a
line is found in a given burst. When necessary, this dependence on $f$
is removed by the Bayesian process of ``marginalization.'' Since
bursts are presumably independent events, $p(D \,|\,fHI)$ is the
product of the probabilities of obtaining the observed detections or
nondetections in each burst.  Thus, for the $N_B$ BATSE bursts in
which no lines have been detected 
\begin{equation}
p(D \,|\,fHI) = \prod_{\sigma = 1}^{N_B} \left(1-p\left(L_\sigma \,|\, 
  f HI\right)\right)
\end{equation}
where $L_\sigma$ is the proposition that a line has been detected in
the $\sigma$th burst, and therefore $p(L_\sigma \,|\, fHI)$ is the
probability of detecting a line in this burst.  If present, a line may
persist over a time range shorter than the burst duration, and will be
found in the $i$th spectrum accumulated during the burst.  Therefore
$p(L_\sigma \,|\, fHI)$ will be a function of the probability $p(L_i
\,|\, fHI)$ of detecting the line in the $i$th spectrum; $L_i$ is the
proposition that the line was detected in the $i$th spectrum.  The
connection between $p(L_\sigma \,|\, fHI)$ and $p(L_i \,|\, fHI)$ is
presented in this paper.  The line detection may be real or false, and
therefore (Band et al. 1995, hereafter Paper~III) 
\begin{equation}
p(L_i \,|\, fHI) = p(L_i \,|\, l_i HI)p(l_i \,|\, fHI)+
p(L_i \,|\, \bar l_i HI)p(\bar l_i \,|\, fHI) \quad ,
\end{equation}
where $l_i$ is the proposition that the line is present in the $l_i$. 
Thus $p(L_i \,|\, l_i HI)$ is the probability of detecting a line in a
spectrum and $p(L_i \,|\, \bar l_i HI)$ is the probability of a
``false positive.'' Currently we assume that our detection criteria
are stringent enough to make the false positive rate negligible, an
assumption we will investigate in the future.  In this paper we also
discuss the database describing the BATSE spectra necessary to
calculate $p(L_i \,|\, l_i fHI)$; we also present some results of
utilizing this database. 

The probability $p(L_i \,|\, l_i HI)$ of detecting a line which is
present in the $i$th spectrum is the foundation of our methodology. 
If they exist, lines are undoubtedly characterized by a currently
unknown distribution of energy centroids, line widths, intensities
(e.g., equivalent widths), and perhaps other parameters. The
detectability of each line type can be considered separately; we
generally have been using the lines reported by \Ginga\ in the S1
segment of GB~870303 (Graziani et al. 1992) and in GB~880205 (Murakami
et al. 1988) as archetypes, although a generic set of lines can be
used.  We find (Paper~III) that the detectability of a given line is a
function of the strength of the continuum (i.e., the signal-to-noise
ratio---SNR) and the angle between the detector normal and the burst
(the burst angle).  This study of line detectability shows that BATSE
would have detected the lines reported by \Ginga, assuming of course
that BATSE functions as modeled. 

Next we need $p(L_{\sigma} \,|\, l_{\sigma} HI)$, the probability of
detecting a line somewhere in the $\sigma$th burst given that the line
is indeed present, which requires a relationship between the presence
of a line in the $\sigma$th burst and its presence in the $i$th
spectrum within this burst.  Thus we may calculate that there is a
very high probability of detecting a line in a given spectrum if it is
present ($p(L_i\,|\,l_i HI)\simeq 1$), but may conclude that even if a
line is present somewhere in the burst, it is most likely not in the
spectrum in question ($p(l_i\,|\,l_\sigma HI)\ll 1$). For example,
based on empirical evidence or theoretical prejudice we may believe
that lines do not persist for the length of time over which the
spectrum was accumulated.  The few line detections from all the burst
missions are insufficient to map out the line distribution (e.g.,
energies, intensities, widths, persistence times), and therefore we
must model the probability $p(l_i \,|\, l_\sigma H)$ that a line
occurs in the $i$th spectrum, regardless of whether it is detectable. 

With $p(L_\sigma \,|\, l_\sigma HI)$ for the bursts observed by
different missions we can evaluate the consistency between these
missions.  Both BATSE and \Ginga\ provided sufficient data to carry
out such an analysis. We developed a number of measures of the
consistency between these two missions using both standard
``frequentist'' (Paper~I) and Bayesian statistics (Paper~II). In
addition, values of $p(L_\sigma \,|\, l_\sigma HI)$ for the bursts in
the BATSE database can be used to identify the most promising bursts
for further analysis.  We first apply new line search techniques
(Schaefer et al. 1994; Briggs et al. 1996) to bursts in which lines
are most likely to be detected. 

The statistical analysis outlined above requires a characterization of
all the spectra from BATSE and \Ginga, and measures of line
detectability for both instruments.  Paper~III provides the line
detection probabilities for the BATSE Spectroscopy Detectors (SDs),
the relevant BATSE detectors. Below we describe the database of BATSE
spectra created to characterize the BATSE bursts. Fenimore et al.
(1993) performed a preliminary evaluation of the \Ginga\ data for the
GB~880205 line set; a more extensive extraction of the necessary
\Ginga\ information is planned. 

Here we consider a ``real'' line to be a true feature in the spectrum
that arrives at the detector; most likely a real feature was emitted
by the burst, although the feature may possibly have been imposed on
the spectrum by astrophysical processes between the burst source and
the detector. A ``detection'' is a feature which satisfies the
detection criteria and is therefore considered to be a real line. Note
that we treat the detection of a line as a binary conclusion:  a
feature is either considered to be a detected line or it is ignored in
subsequent analysis.  Unlike frequentist statistics, in which a
hypothesis is either true or false, Bayesian statistics (which is used
in Paper~II) allows our confidence in a hypothesis to be quantified
via a probability that the hypothesis is true.  In principal we could
develop a formalism which permits a fractional detection through the
probability that a feature is real.  We could then develop a
methodology of using the entire spectral database to tease out the
line distribution from all the spectral bumps and wiggles, most of
which are undoubtedly fluctuations, but a small fraction of which
might result from an underlying distribution of real lines.
Specifically, deviations from the distribution of line significances
expected from mere fluctuations could be used to estimate the
distribution of true spectral lines; this assumes that the fluctuation
distribution can be estimated accurately.  However, most scientists
are more comfortable working with definite detections and
nondetections, and that is the route we have taken. 

Our detection criteria are 1) that a spectral feature is significant
in the spectrum from one detector, and 2) that all spectra from the
detectors which observed the burst are consistent.  Until recently
significance was defined using the $F$-test.  However, as discussed in
the appendix, the $F$-test is appropriate when the uncertainties on
the measured quantities---here the counts in each channel---are
unknown (Eadie et al. 1971), which is not the case here. We have
therefore adopted a maximum likelihood ratio test which uses $\Delta
\chi^2$ as the relevant statistic (D.~Lamb, private communication,
1995).  In practice, these two tests usually give comparable results. 

This paper begins with the method for connecting the probabilities of
detecting a line in a given spectrum and anywhere in a burst (\S 2). 
The resulting methodology requires the line detection probability for
every burst spectrum accumulated by BATSE; a database of parameters
describing these spectra is required to calculate these probabilities
(\S 3).  This data is used to rank the BATSE bursts by the maximum
signal-to-noise ratio of any spectrum accumulated during a burst. 
This database is also utilized to find the number of bursts in which
different line types could be detected under the assumption of a small
line occurrence frequency (\S 4); this number is crucial to the study
of consistency between the BATSE and \Ginga\ observations (\S 5). In
the first appendix we discuss the maximum likelihood ratio test which
we have adopted to evaluate the significance of an observed line
feature.  The second appendix lists the large number of symbols used
in this work. 
\section{PROBABILITY A LINE IS PRESENT IN A GIVEN BURST}
Here we consider the probability of detecting a line in a single
burst, and therefore we suppress the (greek) indices specifying the
burst.  Also, we consider the probability of detecting a line of a
given type specified by parameters such as energy centroid, intrinsic
width and intensity (but not the time over which the line is present);
the resulting calculation must be done for each line type. BATSE
accumulates a series of consecutive spectra from four different SDs
(the Spectroscopy detector High Energy Resolution,
Burst---SHERB---data type); we refer to these basic spectra which
provide the finest time resolution available as SHERB spectra.  We
assume there are $N$ SHERB spectra for a given detector across the
burst from which it is possible to construct $N(N+1)/2$ different
averaged spectra composed of consecutive SHERB spectra; quantities
describing these averaged spectra are specified by roman indices
(e.g., $l_i$ means a line exists in the $i$th averaged spectrum). 

The probability of detecting a line in the burst as a whole is the
probability of detecting a line in at least one of the spectra which
can be searched, 
\begin{equation}
p(L \,|\, fHI) = 1-\prod_{i=1}^{N(N+1)/2} 
   \left[1-p\left(L_i \,|\, fHI \right)\right]
\end{equation}
(the probability of at least one detection is one minus the
probability of no detections).  As will be discussed below, the line
frequency $f$ is the probability that a line is present somewhere in
the burst, $f=p(l \,|\,HI)$. A line detection in a given spectrum
results either from the detection of a real line or a spurious
detection (i.e., a ``false positive''), 
\begin{eqnarray}
p(L_i \,|\, fHI) &=& p(L_i \,|\, l_i HI) p(l_i \,|\, fHI) +
   p(L_i \,|\, \bar l_i HI) p(\bar l_i \,|\, fHI) \quad , \nonumber \\
&=& p(L_i \,|\, l_i HI) p(l_i \,|\, fHI) +
   p(L_i \,|\, \bar l_i HI) \left(1-p(l_i \,|\, fHI)\right) \quad ,
\end{eqnarray}
where we use the fact that $p(l_i \,|\, fHI)$ and $p(\bar l_i \,|\,
fHI)$ are exhaustive.  Paper~III calculated $p(L_i \,|\, l_i HI)$ for
the BATSE SDs, while the probability of a spurious detection $p(L_i
\,|\, \bar l_i HI)$ will be studied further, but is clearly dependent
on the detection threshold. 

Our focus here is $p(l_i \,|\, fHI)$, which is a statement of how
lines occur in burst spectra.  Is the probability that a line occurs
in a burst the same for all bursts, or does it depend on duration,
spectral hardness or other burst properties?  Do lines persist for a
long time or for short intervals? Unfortunately, since there have been
very few detections, we know very little about $p(l_i \,|\, fHI)$. 
Therefore, we have to construct reasonable models of $p(l_i \,|\,
fHI)$ which we will use for further calculations. 

Let $dp(l\,|\,t_b\,t_e\,fHI) = g(t_b,t_e) dt_b dt_e$ be the
probability density for a line beginning at $t_b$ and ending at $t_e$.
 If we assume that the probability depends only on the time a line
persists, and does not favor the beginning or end of the burst, then
$g(t_b,t_e)$ will depend only on the persistence time $t_e-t_b$, i.e.,
$g(t_b,t_e)=g(t_e-t_b)$. Since the data consist of discrete spectra,
we cannot isolate the spectrum over the precise interval during which
a line is present (if such exists since the line intensity may vary). 
Instead, the line will be attributed to a particular sum of
consecutive SHERB spectra with an accumulation period overlapping the
time the line was actually present; conversely, a spectrum summed from
a number of SHERB spectra may show lines with a variety of beginning
and end times. The probability that a line begins between $t_{b1}$ and
$t_{b2}$ and ends between $t_{e1}$ and $t_{e2}$, and would be
attributed to the $i$th spectrum accumulated between $t_j$ and $t_k$
($t_{b1} \le t_j \le t_{b2} \le t_{e1} \le t_k \le t_{e2}$), is 
\begin{equation}
p(l_i \,|\, fHI) = 1-\exp\left[-\int^{t_{b2}}_{t_{b1}} dt_b
\int^{t_{e2}}_{t_{e1}} dt_e \, g(t_b,t_e) \right] \quad ;
\end{equation}
the probability of a line existing in the burst is
\begin{equation}
p(l \,|\, fHI) = 1-\exp\left[-\int^{T}_{0} dt_b
\int^{T}_{t_b} dt_e \, g(t_b,t_e) \right] \quad ,
\end{equation}
where $T$ is the burst duration.  These expressions are derived from
$1-p(l \,|\, fHI) = \prod_j 1-p(l_j \,|\, fHI)= \exp[\sum_j
\ln(1-p(l_j \,|\, fHI))] = \exp [-\sum_j p(l_j \,|\, fHI)]$, where
$\ln (1-p(l_j \,|\, fHI))= -p(l_j \,|\, fHI)$ is valid because $p(l_j
\,|\, fHI)=g(t_b,t_e)dt_b dt_e$ (i.e., $p(l_j \,|\, fHI)$ is small). 
Finally, $\sum_j p(l_j \,|\, fHI) = \int g(t_b,t_e)dt_b dt_e$. In
practice, if the $i$th spectrum begins at $t_j$ and ends at $t_k$, we
use $t_{b1} = (t_{j-1} + t_j)/2$, $t_{b2} = (t_j + t_{j+1})/2$,
$t_{e1} = (t_{k-1} + t_k)/2$ and $t_{e2} = (t_k + t_{k+1})/2$, with a
somewhat more complicated expression for a single SHERB spectrum. 

As examples, we consider three different functional forms for
$g(t_e-t_b)$.  In each case there are two major variants.  The first
variant (eqs. [7], [9], and [11] below) assumes that $g$ is the same
function of the persistence time $t_e-t_b$ with the same normalization
for all bursts, and thus the line frequency varies from burst to burst
(i.e., lines are more likely to occur in long bursts). The second
variant (eqs. [8], [10], and [12] below) assumes the line frequency is
the same for all bursts and therefore the normalization of $g$ varies
from burst to burst.  In practice we use the second case. 

{\it Model 1:}  $g(t_e-t_b)=c$.  If $c$ is the same for all bursts
then the first variant of this model is 
\begin{eqnarray} 
p(l_i \,|\, cHI) &=& 1-\exp \left[- c\left(t_{b2}-t_{b1} \right)
   \left(t_{e2}-t_{e1}\right) \right] \quad ,\nonumber\\
p(l \,|\, cHI) &=& f = 1-\exp\left[-cT^2/2\right] \quad ,
\end{eqnarray} 
where $0\le c\le \infty$.  Note that $f$ increases with the duration
$T$. On the other hand, if $p(l \,|\, cI)=f$ for each burst then the
second variant is 
\begin{equation} 
p(l_i \,|\, fHI) = 1- (1-f)^{2(t_{b2}-t_{b1})(t_{e2}-t_{e1})/T^2} \quad ,
\end{equation} 
where $0\le f\le 1$.  Note that there is no dependence on the
persistence time $t_e-t_b$. 

{\it Model 2:}  $g(t_e-t_b)=cb^2 \exp[-b(t_e-t_b)]$.  This model would
result from a sequence of independent line occurrences, that is, the
probability of the line occurring in any given time interval does not
affect the probability of its presence in the next time interval.  If
$c$ is constant for all bursts then 
\begin{eqnarray} 
p(l_i \,|\, cHI) &=& 1-\exp\left[c \left(e^{bt_{b2}}-e^{bt_{b1}}\right)
   \left (e^{-bt_{e2}}-e^{-bt_{e1}}\right)\right] \quad ,\nonumber\\
p(l \,|\, cHI) &=& f = 1-\exp\left[-c\left (e^{-bT}
   -\left(1-bT\right) \right) \right] \quad ,
\end{eqnarray} 
where $0\le c\le \infty$.  If we assume $p(l \,|\, cHI)=f$ then
\begin{equation} 
p(l_i \,|\, fHI) = 1- \left(1-f\right)^{ 
   \left(e^{bt_{b2}}-e^{bt_{b1}}\right)
   \left(e^{-bt_{e2}}-e^{-bt_{e1}}\right)/
   \left(\left(1-bT\right) -e^{-bT}  \right)} \quad ,
\end{equation} 
where $0\le f\le 1$.  In this case there is a strong dependence on the
persistence time $t_e-t_b$ since $\left(e^{bt_{b2}}-e^{bt_{b1}}\right)
   \left(e^{-bt_{e2}}-e^{-bt_{e1}}\right)= 
   \exp\left[-b\left(t_{e2}-t_{b2}\right)\right]
   \left(1-e^{-b(t_{b2}-t_{b1})}\right)
   \left(1-e^{b(t_{e2}-t_{e1})}\right)$.

{\it Model 3:}  $g(t_e-t_b) = c (t_e-t_b)^{-b}$.  Assuming $c$ is
constant for all bursts 
\begin{eqnarray}
p(l_i \,|\, cHI) &=& 1-\exp\left[-{{c\left( \left(t_{e2}
   -t_{b1}\right)^{2-b}-\left(t_{e2}-t_{b2}\right)^{2-b}
   -\left(t_{e1}-t_{b1}\right)^{2-b}+\left(t_{e1}-
   t_{b2}\right)^{2-b} \right)}\over{\left(2-b\right)
   \left(1-b\right)}} \right] , \nonumber\\
p(l \,|\, cHI) &=& f = 1-\exp[-cT^{2-b}/(2-b)(1-b)] \quad ,
\end{eqnarray}
where once again $0\le c\le \infty$.  Requiring $p(l \,|\, cHI)=f$ for
all bursts leads to 
\begin{equation}
p(l_i \,|\, fHI) = 1- (1-f)^{\left( \left(t_{e2}-t_{b1}\right)^{2-b}
-\left(t_{e2}-t_{b2}\right)^{2-b} -\left(t_{e1}-t_{b1}\right)^{2-b}
+\left(t_{e1}-t_{b2}\right)^{2-b} \right)/T^{2-b}} \quad ,
\end{equation}
where as usual $0\le f\le 1$.  Here there is also a strong dependence
on the persistence time $t_e-t_b$. 
\section{DATABASE}
In Paper~III we found that the probability $p(L_i \,|\,l_i HI)$ of
detecting a line in a spectrum was a function of the spectrum's SNR
and the burst angle. Therefore we need these quantities for each
spectrum from all the detectors for which there are data for a given
burst.  Because \Ginga\ reported lines at $\sim 20$~keV and $\sim
40$~keV, we use SNR calculated between 25 and 35~keV. Thus the SNR
measures the strength of the continuum in the energy range of interest
which should mitigate the effect of different shape continua. 

We would like to search spectra with arbitrary beginning and end
times, but the telemetry only provides spectra with discrete beginning
and end times.  Our search is meant to find the combination of
consecutive SHERB spectra in which a candidate feature has the
greatest significance.  Thus, if $N$ SHERB spectra span a burst, we
need to consider $N(N+1)/2$ spectra. However, the database does not
need to store parameters for all $N(N+1)/2$ possible spectra since
they can be calculated from a smaller set of data.  Here we assume
that the burst angle and background count rate $R_B$ are constant for
the entire burst for the detector providing the SHERB spectra, a
reasonable assumption since the burst durations are usually less than
$100$~s (the time scale over which the background rate might change
significantly enough to affect our results; the burst angle will
change on much longer time scales). The SNR for each possible spectrum
can be calculated from the counts and accumulation time for each SHERB
spectrum. Thus 
\begin{equation}
SNR_i = {{C_i-R_B \Delta t_i}\over \sqrt{C_i\Delta E}}
\end{equation}
where $C_i$ is the number of detected counts summed over all the SHERB
spectra of which the $i$th spectrum consists, $R_B$ is the background
count rate, $\Delta t_i$ is the time over which the spectrum was
accumulated (i.e., the sum of the accumulation times of the
constituent SHERB spectra), and $\Delta E$ is the size of the energy
range ($\Delta E \sim 10$ keV).  Both $R_B$ and $C_i$ are accumulated
over $\Delta E$. The factor of $\Delta E^{-1/2}$ converts the SNR from
a ratio using the counts over an energy range ($\Delta E$ will vary in
size from detector to detector and burst to burst) to a ratio using
the counts per keV. Note that a livetime correction is not made.  Thus
the database need contain parameters only for each SHERB spectrum, as
discussed in detail below. 

For a burst to be included in our database it had to have a peak count
rate in the Large Area Detectors (LADs) over $\sim$7,500 s$^{-1}$ in
the 50--300~keV energy band; of the 1550 bursts on which BATSE
triggered between 1991 April and 1996 May, 297 met this criterion.
After identifying the channels between 25 and 35~keV, we extracted the
number of counts in these channels for each SHERB spectrum for all the
detectors that provided data.  The background counting rate is the
time average from a series of SHERB spectra after the burst, if
available, and SHER spectra (Spectroscopy detector High Energy
Resolution---background spectra accumulated when BATSE is not in burst
mode) before and after the burst, if necessary. Calculating higher
accuracy backgrounds is unnecessary for our purposes since here we
only need a measure of the strength of the burst, not an accurate
background-subtracted spectrum for spectral fitting.  In some cases
the calculated background was clearly too high---indicated by a large
number of negative background-subtracted count rates---or low---found
by inspecting weak bursts with large SNRs.  Incorrect background rates
were recalculated, often using stretches of background in the middle
of, or just after, a burst.  A burst for which the SNR is sensitive to
the background level is usually too weak to harbor detectable spectral
lines. Spectra from all detectors were included in the database, but
we ignored data from detectors set at low gain or with burst angles
greater than $\sim85\arcdeg$:  low gain detectors have a low energy
cutoff $E_{\rm low}$ above the energies at which lines have been
observed, and the spacecraft shields the detectors for very large
burst angles. Line detectability depends on the energy range covered;
a line at 20~keV will not be detected if the spectrum begins at
20~keV.  Therefore we also characterized each spectrum by its low
energy edge $E_{\rm low}$, which we define as the upper end of the
SLED (an electronic artifact just above a spectrum's true low energy
cutoff---Band et al. 1992). The database therefore consists of the
following data for each detector for each burst: the time interval
over which the SHERB spectra were accumulated; the number of counts in
the 25--35~keV range for each SHERB spectrum; and additional
information for each burst-detector pair such as the burst angle, the
energy $E_{\rm low}$ of the upper end of the SLED, and the exact
energy width $\Delta E$ of the 25--35~keV range. 

The product of the methodology and database described above is the
probability for each burst of detecting a line if present $P(L\,|\,l
HI)$. The primary purpose of this probability is to assess the
consistency between BATSE and other missions, and to estimate the
frequency with which lines occur.  However, this probability can also
be used to identify the bursts in which lines are most detectable. Our
search should therefore focus on those bursts. As an example, we
characterized each burst by the maximum SNR for any spectrum during
the burst.  Figure~1 presents the cumulative distribution.  Since the
gain, and thus the energy range included in the spectrum, varies from
burst to burst and detector to detector, we show distributions by
maximum $E_{\rm low}$. Thus a line at 20~keV would be detectable in
those bursts with a detector for which $E_{\rm low}<15$~keV. From
Paper~III we find that the GB~880205 line set (19.4 and 38.8~keV)
would have been detected half the time by BATSE for SNR$\simeq 7$. The
line at 38.8~keV appears to determine the detectability of this line
set in the BATSE spectra, and therefore we require $E_{\rm low}\le
25$~keV.  We see that the line at $\sim40$~keV would have been
detectable in the highest SNR spectrum in about 65 bursts. 
\section{SIMPLIFIED LIKELIHOOD CALCULATION}
The Bayesian consistency measures and related quantities require $p(D
\,|\,fHI)$, the probability of obtaining the observed results $D$
assuming a hypothesis $H$ about bursts and the detectors (Paper~II);
$p(D \,|\,fHI)$ is also known as the likelihood for $f$ and $H$.  Thus
$D$ might represent the absence of a BATSE line detection or the
\Ginga\ line detections in specific bursts, while $H$ might stand for
the hypothesis that lines exist, the BATSE detectors are modeled
correctly, and the BATSE and \Ginga\ results are consistent.  In our
formulation we explicitly separate out the line frequency $f$. A burst
with a detection contributes to $p(D \,|\, fHI)$ a factor of
$p(L_\sigma \,|\,fHI)$, while a burst with no line detection
contributes $1-p(L_\sigma \,|\,fHI)$.  Note that as before roman and
greek indices specify spectra and bursts, respectively.  Thus, if
there are line detections in $n_d$ bursts in a database of $N_B$
bursts then 
\begin{equation}
p(D \,|\, fHI) = \prod_{\sigma=1}^{N_B} \left(1-p \left(L_\sigma \,|\, 
   fI\right)\right) \prod_{\rho=1}^{n_d} {{p \left(L_\rho \,|\, fI\right)}
   \over {\left(1-p \left(L_\rho \,|\, fI\right)\right)}}
\end{equation}
where the detections have been placed at the beginning of the
database.  The line frequency $f$ is not a quantity of interest to the
consistency issue, and therefore it is ``marginalized,'' 
\begin{equation}
p(D \,|\, HI) = \int df p(f \,|\, HI) p(D \,|\, fHI) \quad .
\end{equation}
The ``prior'' for $f$, $p(f \,|\, HI)$, is our assessment of the
likely value of $f$ before the data $D$ were obtained.  In general we
assume that $f$ could be any value between 0 and 1, and therefore $p(f
\,|\, HI)=1$. 

We saw in \S 2 that $p(l_i \,|\, fHI)=1-(1-f)^{\gamma_i}$ (e.g., eqs.
[8], [10] or [12]).  Consequently the probabilities of detecting a
line in a given burst $p(L_\sigma \,|\, fHI)$ and of obtaining the
observed database $p(D \,|\, fHI)$ are complicated functions of $f$. 
Thus the integral over $f$ in eq.~(15) will be a time-consuming
numerical calculation since information from all the bursts must be
included in evaluating the integrand at each value of $f$.  However,
we can make some simplifying assumptions.  First, we assume the false
positive probability $p(L_i \,|\, \bar l_i HI)$ is very small and can
be neglected. Second, the absence of a detection in the BATSE dataset
indicates that the line frequency $f$ is probably small, and therefore
\begin{equation}
p(l_i \,|\, fHI) = 1-\left(1-f\right)^{\gamma_i} \simeq \gamma_i f \quad .
\end{equation}
Consequently:
\begin{eqnarray}
p(L_i \,|\, fHI) &=& p(L_i \,|\, l_i HI) \gamma_i f \quad , \\
p(L_\sigma \,|\, fHI) &=& \left[ \sum_{i=1}^{N_\sigma(N_\sigma+1)/2} 
p(L_i \,|\, l_i HI) \gamma_i \right] f \quad , \\
p(D \,|\, fHI) &=& \left(1-\left[ \sum_{\sigma=1}^{N_B} 
   \sum_{i=1}^{N_\sigma(N_\sigma+1)/2} p(L_i \,|\, l_i HI)\gamma_i \right] 
   f \right) \quad , \\
M(L_D \,|\, l_D HI) &\equiv& \sum_{\sigma=1}^{N_B} 
   \sum_{i=1}^{N_\sigma(N_\sigma+1)/2} p(L_i \,|\, l_i HI)\gamma_i \quad .
\end{eqnarray}
$N_\sigma$ is the number of SHERB spectra spanning the $\sigma$th
burst. We approximated $p(D \,|\, fHI)$ in eq.~(19) for the case of no
detections, which is currently relevant for BATSE (i.e., $n_d=0$ in
eq.~[14]). The quantity $M(L_D \,|\, l_D HI)$ is the sum of each
burst's detection probability; thus $M(L_D \,|\, l_D HI)$ will be
nearly equal to the number of bursts if all bursts are uniformly
strong, whereas weak bursts will not contribute to this statistic. 
Since $M(L_D \,|\, l_D HI)$ is the first order expansion in $f$, it is
valid for small values of $f$, i.e., under the assumption that a line
is unlikely to be present in a given burst.  The small $f$
approximation in eq.~(19) is valid only for $f\ll 1/M(L_D \,|\, l_D
HI)$; note that $(1-f)^m \simeq 1-mf$ is not accurate for $f \ge 1/m$
even if $1/m$ is small.  However we shall use this approximation to
$f\sim 1/M(L_D \,|\, l_D HI)$. We can now marginalize $f$ to obtain 
\begin{equation}
p(D\,|\,HI) = \int df p(f \,|\, HI) p(D\,|\, fHI) = {1\over 
   {2M(L_D\,|\, l_D HI)}} \quad ,
\end{equation}
where we set $p(D\,|\, fHI)= 0$ for $f\ge 1/M(L_D \,|\, l_D HI)$.
Using the expression in Paper~II for the probability distribution for
$f$ given the new data $D$ (here the absence of a BATSE line
detection) we find 
\begin{eqnarray}
p(f \,|\, DHI) &=& {{p(f \,|\,HI)p(D\,|\,fHI) }\over{p(D\,|\,HI) }} \\
   &=& \cases{2M(L_D\,|\, l_D HI) 
   \left(1-M\left(L_D\,|\, l_D HI\right)f\right) 
   \quad , \quad & $f \le 1/M(L_D\,|\, l_D HI)$ \cr
   0 \quad , \quad & $f \ge 1/M(L_D\,|\, l_D HI)$ \cr} \nonumber
\end{eqnarray}
where we used the prior $p(f \,|\,HI)=1$.  In both eqs. (21) and (22)
we extend the approximation in eq.~(16) to the regime $f\simeq 1/M(L_D
\,|\, l_D HI)$ where the approximation will have broken down.  In
Figure~2 we compare $p(f\,|\,DHI)$ in eq.~(22) to $p(f \,|\, DHI)=
(m+1)(1-f)^m$ which results from the absence of a line detection in
$m$ bursts in which lines could have been detected with 100\%
probability (i.e., $p(L_\sigma \,|\, l_\sigma HI)=1$).  As can be
seen, the small $f$ approximation is accurate to a factor of $\sim2$
in normalization and extent.  Given the uncertainties and other
approximations in this analysis, this accuracy is sufficient.  In
part, the small $f$ approximation demonstrates the utility of
$M(L_D\,|\, l_D HI)$ as a diagnostic statistic. 
\section{DISCUSSION}
The quantity $M(L_D\,|\, l_D HI)$ characterizes the detectability of
spectral lines in a burst database and thus our ability to learn about
lines from the database.  Primarily, $M(L_D\,|\, l_D HI)$ is the
approximate number of bursts in which lines could be detected.  We
have seen that its inverse is twice $p(D\,|\,HI)$, the likelihood of
the hypothesis $H$ and that it is the width of the distribution for
the line frequency $f$.  Using the burst database described in \S 3 we
calculated $M(L_D\,|\,l_D HI)$ for the BATSE spectra. Since the burst
distribution---the frequency of lines of different types and where
within the burst they occur---is unknown, we made a number of modeling
assumptions.  These assumptions, along with the supposition that lines
exist and that the modeling of the BATSE detectors is correct,
constitute the hypothesis $H$.  First, we assume that the line
frequency $f$ is the same for all bursts, and that a line can occur in
any spectrum with equal probability (the second variant of model~1 in
\S 2).  Thus we use eq.~(8) to define $\gamma_i$, i.e.,
$\gamma_i=2(t_{b2}-t_{b1})(t_{e2}-t_{e1})/T^2$.  Second, we
approximate $p(L_i \,|\, l_i HI)$, the probability of detecting a line
in a spectrum, as 1 for SNRs above a threshold value if the low energy
cutoff is less than a certain energy.  In Paper~III we found that this
probability $p(L_i\,|\,l_i HI)$ was a function of both the SNR and the
burst angle, and that the transition from 0 to 1 occurs over a range
of SNR. The calculations in Paper~III assumed a low energy cutoff of
10~keV which is rarely achieved because of the SLED electronic
artifact which raises the effective cutoff (Band et al. 1992) and the
gain settings of the SDs. However, in Band et al. (1996, henceforth
Paper~IV) we showed that the detectability of a line is insensitive to
the low energy cutoff as long as sufficient continuum is included
below the line candidate (in the example there 15--20~keV). We have
been using the two \Ginga\ detections to characterize the unknown line
distribution. For the line at 21.1~keV in the S1 segment of GB~870303
the transition between a detection probability of 0 and 1 occurs at a
SNR of $\sim2$, while for the harmonic lines in GB~880205 at 19.4 and
38.8~keV the transition occurs at a SNR of $\sim 7$; in both cases the
detectability is also angle dependent.  However, for greater
generality we present in Figure~4 $M(L_D\,|\, l_D HI)$ for a range of
SNRs and low energy cutoffs.  As can be seen, $M(L_D\,|\, l_D
HI)\simeq 50$ for detecting lines similar to the one in GB~870303,
assuming a low energy cutoff of 15~keV will suffice. If the
detectability of the GB~880205 lines is dominated by the 38.6~keV line
(see Figure~1 of Paper~III), and thus a low energy cutoff less than
25~keV is necessary, then $M(L_D\,|\, l_D HI)\simeq 20$.  It is clear
from these curves that despite the large number of bursts observed by
BATSE in the past 5 years, lines would be detectable in relatively few
bursts. 

Only for the rare strong bursts are lines detectable in spectra
accumulated by \Ginga\ and the BATSE SDs.  Since BATSE detects bursts
with a much larger detector than the SDs, whereas \Ginga\ detected
bursts with the same detector which accumulated spectra, the BATSE
burst database includes a larger fraction of weak bursts. In addition,
\Ginga\ reported lines at $\sim20$ and $\sim40$~keV. The \Ginga\ burst
detector was sensitive down to 2~keV (Murakami et al. 1989) whereas
BATSE's $E_{\rm low}$ is typically $\sim$20~keV.  On the other hand,
each BATSE SD has an area twice that of the \Ginga\ detector.
Therefore BATSE is usually less sensitive than \Ginga\ to lines below
$\sim40$~keV, and more sensitive above 40~keV. 

The necessary \Ginga\ data has not yet been extracted to complete the
study of the consistency between the BATSE and \Ginga\ observations as
presented in Paper~II.  However, the small values of $M(L_D\,|\, l_D
HI)$ for the BATSE bursts and the preliminary value of $M(L_D\,|\, l_D
HI)\simeq 5.4$ for a \Ginga\ detection of lines similar to the one in
GB~880205 (Fenimore et al. 1993), indicate that the apparent
discrepancy between BATSE and \Ginga\ is not severe.  For example, the
probability that the one detection of a line similar to GB~880205
occurred in the \Ginga\ bursts and not in the BATSE data is (Papers I
and~II) $P\simeq M_{\Ginga}/ (M_{\Ginga}+M_{\rm BATSE})\sim
5.4/(5.4+20) \sim 0.2$, which is hardly an improbable event. 
\section{SUMMARY}
We now have a methodology which provides the probability for detecting
a line in the bursts observed by BATSE.  This probability can be used
to evaluate the consistency between the line detections and
nondetections by BATSE and other burst missions, estimate the
frequency with which lines occur, and identify bursts in which lines
are likely to be discovered by new search techniques.  The new element
in the methodology is the weighting of the probabilities for detecting
a line in each of the spectra spanning the burst which can be formed
from the SHERB spectra provided by the telemetry. This weighting is
model-dependent; we explored three models where the line occurrence
depends only on the time a line persists. 

Implementing this methodology requires parameter values characterizing
each spectrum in the bursts under consideration.  To this end we built
a database of the necessary information.  We have been using this
database to identify the bursts in which lines may be detected. 

We calculated the number of bursts in which lines of various types, as
parameterized by the minimum SNR and maximum low energy cutoff
necessary for a detection, could have been detected if the lines were
indeed present.  This calculation assumes a small line frequency. 
These quantities are necessary for the probability that no lines would
be detected in the BATSE data and for the distribution for the line
frequency, and therefore these numbers are essential for measures of
the consistency between the BATSE and \Ginga\ line observations. 
\Ginga-like lines can be detected in relatively few of the large
number of bursts BATSE has observed; for example, lines similar to the
GB~880205 pair of lines are detectable in only $\sim20$ BATSE bursts.
Although comparable \Ginga\ data is not yet available, the discrepancy
between \Ginga\ and BATSE does not appear to be severe.  For example,
a simple calculation shows that the probability that the GB~880205
line set would be detected in a \Ginga\ burst is 20\%, which is hardly
improbable. 

Finally, to evaluate the significance of line candidates we have
adopted the maximum likelihood ratio test which is more appropriate
than the $F$-test. The $F$-test should be used when the uncertainties
on the datapoints are unknown. These two tests give similar
significances when the reduced $\chi^2$ is of order unity.  Indeed, we
find little change in the significances given by the two tests for the
line candidates identified by the visual search of BATSE spectra. 
\acknowledgments
We thank C.~Graziani, D.~Lamb, and T.~Loredo for insightful discussion
about statistics. We are grateful for the detailed comments of the
referee, A.~Connors.  The work of the UCSD group is supported by NASA
contract NAS8-36081. 
\appendix
\section{SIGNIFICANCE STATISTIC}
To evaluate the significance of a given spectral feature we have been
using the $F$-test which compares fits with nested models (i.e., one
model is a subset of the other).  Assume that $\chi^2_1$ results from
fitting a spectrum of $N_c$ channels by a continuum model with $r_1$
parameters (thus $\nu_1=N_c-r_1$ degrees-of-freedom), and that
$\chi_2^2$ results from fitting the spectrum by a continuum+line(s)
model with $r_2$ parameters ($\nu_2=N_c-r_2$ degrees-of-freedom).  In
the continuum+line(s) model, the $r_1$ continuum parameters are the
same as for the continuum model; thus an additional $\Delta
\nu=r_2-r_1$ parameters have been added by modeling the line(s).  If
the continuum model is correct and there are actually no lines then
the quantity 
\begin{equation} 
F_0 = {{\chi^2_1 - \chi^2_2}\over {\Delta \nu}} / 
   {{\chi^2_2}\over {\nu_2}}
\end{equation} 
is distributed as $F(\Delta \nu,\nu_2)$.  Consequently $P(F\ge F_0)$
is the probability of finding $F$ larger than or equal to $F_0$ when a
continuum+line(s) model is fit to a count spectrum resulting from a
photon spectrum correctly described by the continuum model.  Thus this
is the probability that the improvement in $\chi^2$ by adding the
additional $\Delta \nu$ line parameters is a fluctuation. 

The $F$-test we have been using is based on a maximum likelihood ratio
test where the uncertainties are unknown.  The original version
defines $\chi^2$ without uncertainties, $S^2 = \sum^{N_c}_{i=1}
(y_i-m_i)^2/N_c$ where $y_i$ is the observed value and $m_i$ is the
model value.  Then 
\begin{equation}
F_1={{S^2_1-S^2_2}\over {S^2_2}} {{\nu_1}\over{\Delta \nu}}
\end{equation}
is distributed as $F(\Delta \nu,\nu_1)$ (Eadie et al. 1971, p. 238). 
Since $N_c$ is large, there is little difference between $\nu_1$ and
$\nu_2$. Both $F_0$ and $F_1$ use ratios of $\chi^2$ and $S^2$,
respectively. Thus the $F$ statistic eliminates the effect of a
systematic multiplicative error in the uncertainties used in $\chi^2$
or of an unknown constant uncertainty on the datapoints in $S^2$. 
This demonstrates why the $F$-test is appropriate for the case where
the uncertainties are not known. 

However, we find that the uncertainties in our data result
predominantly if not exclusively from counting statistics, and
consequently the uncertainties are known.  Thus we can use the
fundamental maximum likelihood ratio test (MLRT) from which the
$F$-test is derived.  This test states that $\Delta
\chi^2=\chi^2_1-\chi^2_2$ is distributed as $\chi^2(\Delta \nu)$ if
the continuum model is sufficient and no lines are present (Eadie et
al. 1971, pp. 230--237).  As with the $F$-test, a small value of
$P(\Delta \chi^2 \ge \Delta \chi^2_0)$ indicates a small probability
that the continuum model alone describes the data.  In Figure~4 we
compare the MLRT to the $F$-test (using the $F_0$ statistic of
eq.~[A1]) for the same values of $\Delta \chi^2$ and different values
of the reduced $\chi^2$, $\chi^2_\nu=\chi^2/\nu$. 

As Figure~4 shows, the two tests give the same values for $\chi^2_\nu$
slightly less than 1, which is not surprising since we expect
$\chi^2_\nu \sim 1$ if our spectral model is correct. A value of
$\chi^2_\nu$ which differs significantly from 1 may result from an
incorrect value for the uncertainties used in $\chi^2$, which the
$F$-test attempts to correct.  However, other factors may cause
$\chi^2_\nu$ to differ from unity, such as an incorrect continuum
model and inaccuracies in the detector response model and the energy
calibration. This is a major reason to favor the MLRT.  To determine
the continuum from which a candidate line deviates, we include all the
spectral data, including continuum far from the line.  However, we do
not know the true continuum shape, which might raise $\chi^2_\nu$. 
Also, the $F$-test depends on the total number of datapoints (the $F$
statistic has a distribution which is a function of the number of
degrees-of-freedom). On the other hand, the MLRT is a function of the
number of added parameters $\Delta \nu$. 

In most cases the MLRT and the $F$-test will lead to the same
conclusion as to whether a feature is significant.  Indeed, evaluating
the line candidates from the visual search of BATSE SD spectra with
the MLRT as opposed to the $F$-test (which was used in Paper~IV) does
not lead to a qualitative difference in significance. Figure~5
compares the probabilities given by these two tests.  As concluded in
Paper~IV, none of the line candidates is significant. 
\section{Notation}
The following is a list of the symbols used in this paper.
\begin{list}{}
\item
$b$, $c$---constants used in modeling $g(t_b,t_e)$.
\item
$C_i$---total counts over energy range $\Delta E$ in the $i$th spectrum.
\item
$\chi^2_i$---the $\chi^2$ statistic for the $i$th spectral fit.
\item
$D$---the observations, specifically whether or not lines were
detected in a burst database. 
\item
$\Delta \chi^2$---the difference in $\chi^2$ between continuum and
continuum+line(s) fits to a spectrum with a candidate line feature 
\item
$\Delta E$---width of the energy range over which the SNR is measured.
\item
$\Delta t_i$---accumulation time of the $i$th spectrum.
\item
$\Delta \nu$---number of parameters added by modeling a line
\item
$E_{\rm low}$---low energy edge of the usable energy range.
\item
$f$---the frequency with which a line type is present in any burst. 
The use of $f$ assumes that each burst, regardless of its
characteristics, has the same probability of hosting the line type. 
\item
$g(t_b,t_e)$---probability that the line is present between $t_b$ and
$t_e$. 
\item
$\gamma_i$---the factor in $p(l_i \,|\, fHI) = \gamma_i f$ in the
small $f$ approximation. 
\item
$H$---hypothesis about the presence of lines and the operation of the
BATSE and/or \Ginga\ detectors. 
\item
$I$---the proposition representing our understanding of the burst
detector and other information known or assumed about the burst. 
\item
$l_\sigma$---the proposition that a line is present in the $\sigma$th
burst. 
\item
$l_i$---the proposition that a line is present in the $i$th spectrum. 
An index specifying the burst is suppressed. 
\item
$L_\sigma$---the proposition that a line is detected in the $\sigma$th
burst. 
\item
$L_i$---the proposition that a line is detected in the $i$th spectrum.
 An index specifying the burst is suppressed. 
\item
$M(L_D \,|\, l_D HI)$---the sum of the probabilities for each burst
that a line would be detected in the burst if the line is present,
calculated in the small $f$ approximation. 
\item
$M_{\Ginga}$, $M_{\rm BATSE}$---value of $M(L_D \,|\, l_D HI)$ for the
\Ginga\ and BATSE burst databases, respectively. 
\item
$n_d$---number of line detections in a burst database
\item
$N$, $N_\sigma$---number of SHERB spectra spanning the $\sigma$th
burst. 
\item
$N_B$---number of bursts in a burst database.
\item
$N_c$---number of channels in a spectrum.
\item
$\nu_i$---number of degrees-of-freedom in the $i$th spectral fit.
\item
$p(D \,|\,fHI)$---the likelihood of $H$ and $f$, the probability of
observing $D$ given hypothesis $H$ and line frequency $f$. 
\item
$p(D \,|\,HI)$---the likelihood of $H$, the probability of observing
$D$ given hypothesis $H$.
\item
$p(f \,|\, HI)$---``prior'' for the line frequency, the probability
distribution for $f$ given $H$ and $I$, but without knowledge of $D$. 
\item
$p(f \,|\, DHI)$---the probability distribution for $f$ based on the
observations $D$ given $H$ and $I$. 
\item
$p(L \,|\, fHI)=p(L_\sigma \,|\, fHI)$---the probability of detecting
a line in the $\sigma$th burst assuming a line frequency $f$. 
\item
$p(L_{\sigma} \,|\, l_{\sigma} HI)$---the probability of detecting a
line somewhere in the $\sigma$th burst given that it is indeed
present. 
\item
$p(L_i \,|\, fHI)$---the probability of detecting a line in the $i$th
spectrum assuming a line frequency $f$. 
\item
$p(L_i\,|\,l_i HI)$---the probability of detecting a line in the $i$th
burst given that it is indeed present. 
\item
$p(L_i\,|\, \bar l_i HI)$---the probability of a false positive, i.e.,
of detecting a line in the $i$th burst when none is present. 
\item
$p(l_i \,|\, fHI)$---the probability that the line is in the $i$th
spectrum assuming a line frequency $f$. 
\item
$p(l_i  \,|\, l_\sigma)$---the probability that if a line is present
in the $\sigma$th burst, it is in the $i$th spectrum of that burst. 
\item
$R_B$---background count rate in a detector over energy range $\Delta
E$. 
\item
$r_i$---number of parameters in the $i$th spectral fit.
\item
$t_b$---time line first becomes apparent.
\item
$t_e$---time line is last apparent.
\item
$T$---burst duration.
\end{list}
\clearpage

\clearpage

\figcaption{Cumulative distribution of bursts by maximum
signal-to-noise ratio (SNR). Each burst is characterized by the
maximum value of the SNR in the range 25--35~keV for any spectrum
formed from consecutive SHERB spectra. The curves are for different
maximum values of a spectrum's effective low energy edge, $E_{\rm
low}$.} 

\figcaption{The distribution for the line frequency $f$, $p(f \,|\,
HI)$.  In the first case (solid curve), the sum of the probabilities
for each burst that a line would be detected if present is $M(L_D
\,|\, l_D HI)=40$.  This case is calculated in the approximation that
$f$ is small, which breaks down for $f\sim 1/M(L_D \,|\, l_D HI)$. In
the second case (dashed curve) there are $m=40$ bursts in which lines
are always detectable.} 

\figcaption{$M(L_D \,|\, l_D HI)$ as a function of the threshold
signal-to-noise ratio (SNR) for different maximum low energy cutoffs. 
A line is detectable in a spectrum if the spectrum's SNR in the
25--35~keV band exceeds the threshold and the low energy cutoff
$E_{\rm low}$ is less than the maximum value labeling each curve.
$M(L_D \,|\, l_D HI)$ is a measure of: the number of bursts in which
lines could have been detected; the inverse of the likelihood; and the
width of the distribution for the line frequency.} 

\figcaption{A comparison between significance given by the maximum
likelihood ratio test (MLRT---solid curve) and the $F$-test (dashed
curves) as a function of $\Delta \chi^2$. A BATSE line candidate
scenario was used: a 4 parameter continuum model, a 3 parameter line
model, and 200 datapoints.  The $F$-test is shown for the labeled
values of the reduced $\chi^2$.} 

\figcaption{Comparison between the significances given by the maximum
likelihood ratio test (MLRT) and the $F$-test for the line candidates
identified by the visual search of BATSE spectra (Band et al. 1996).} 

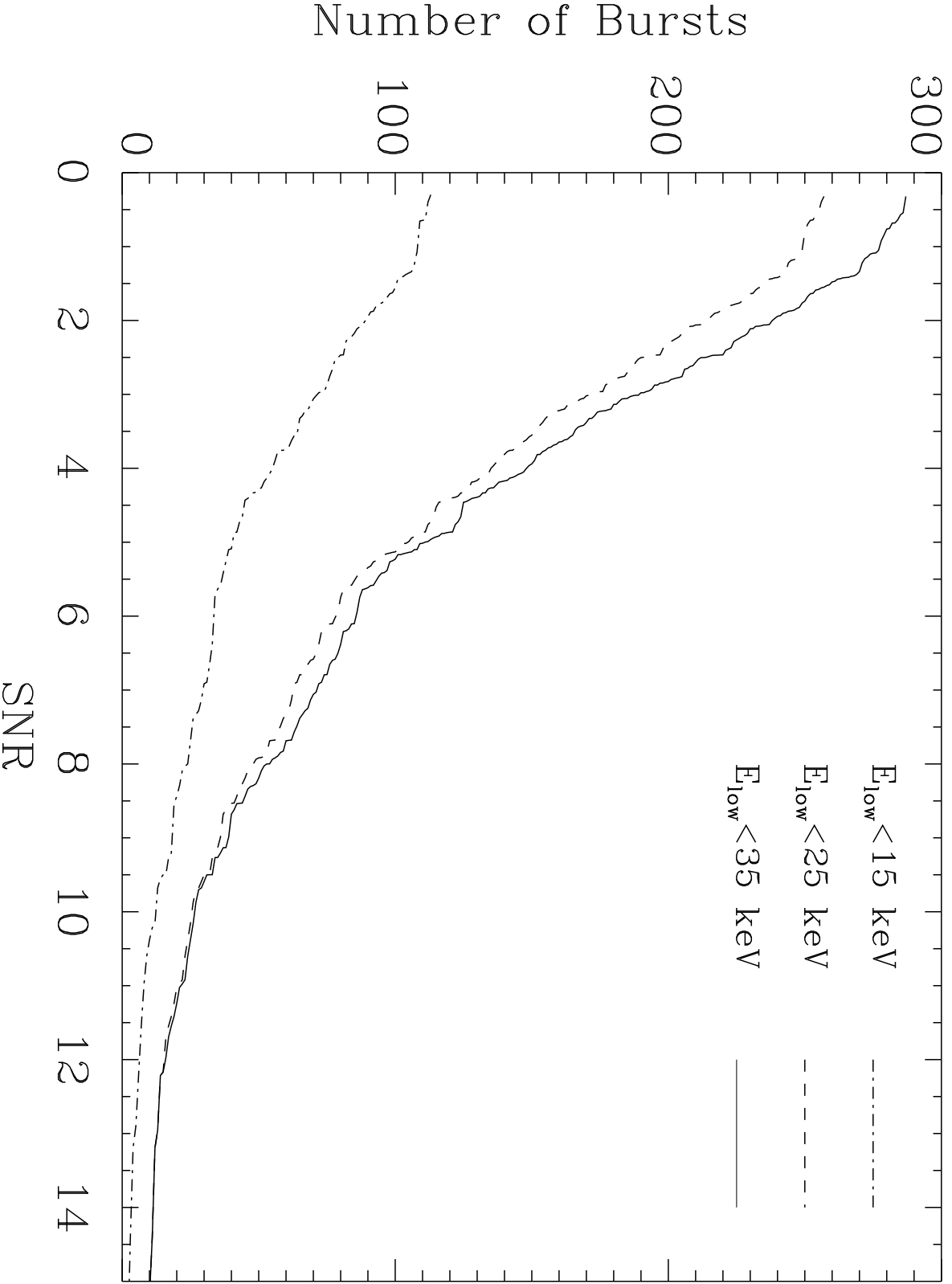
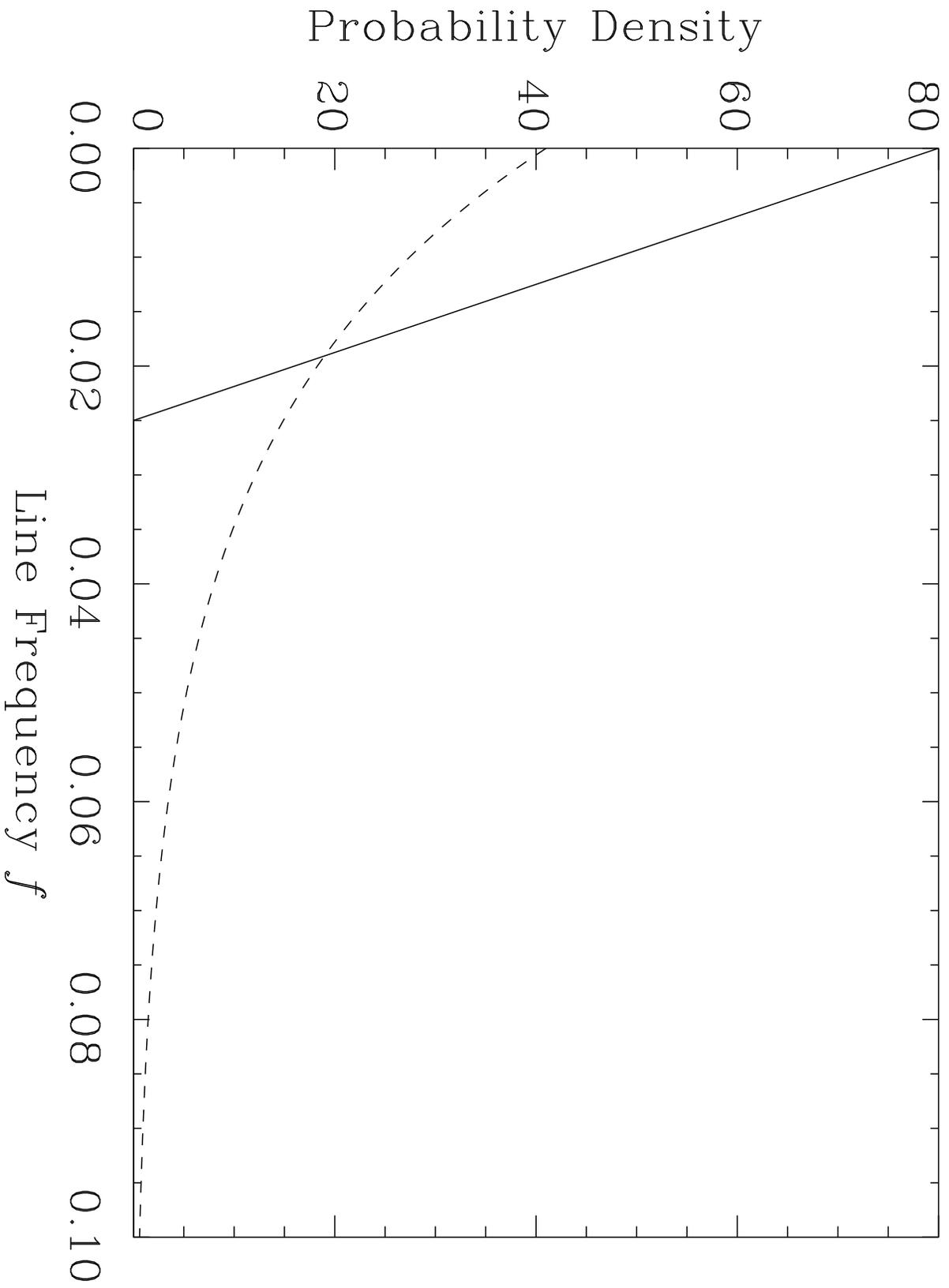
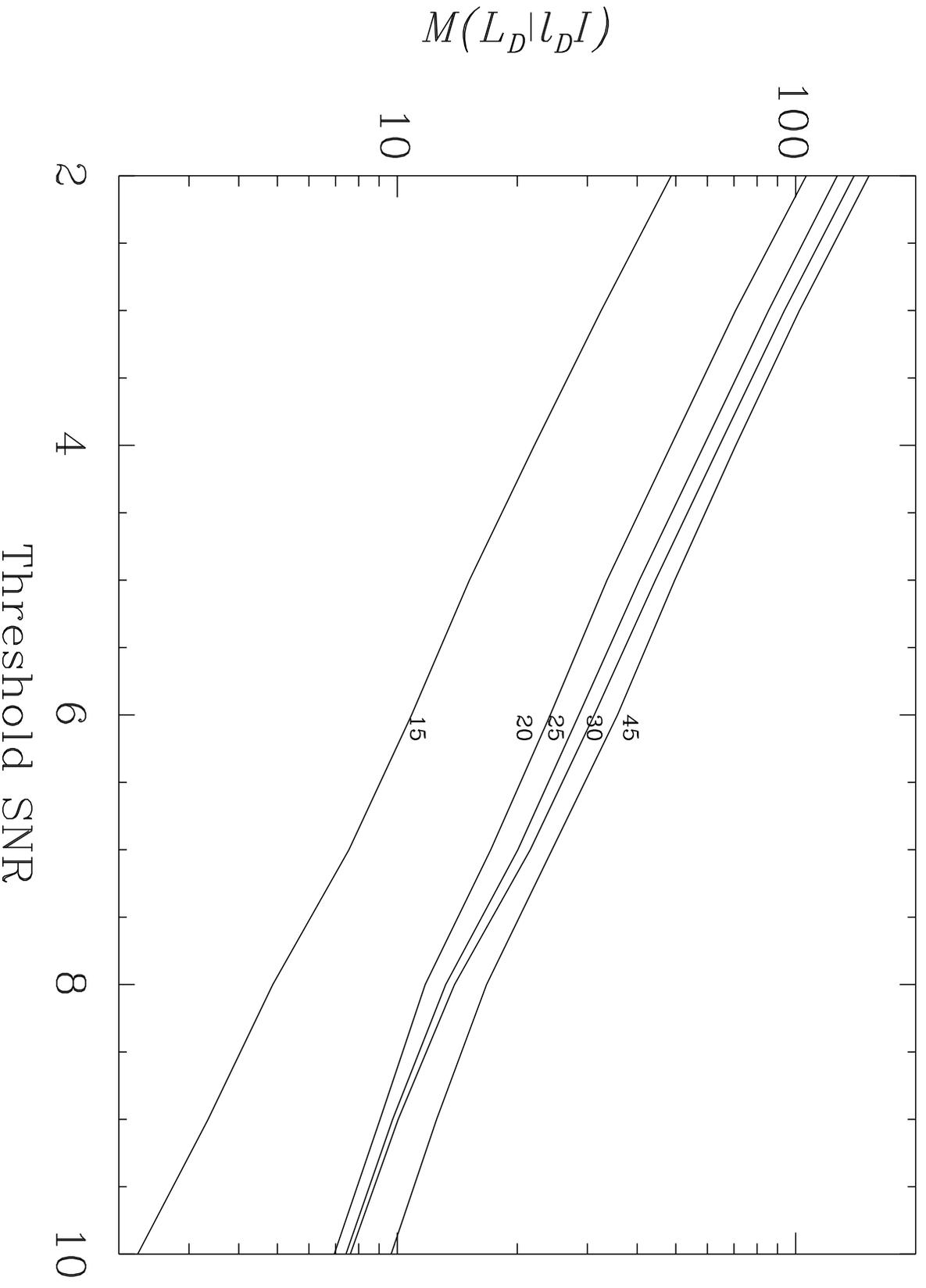
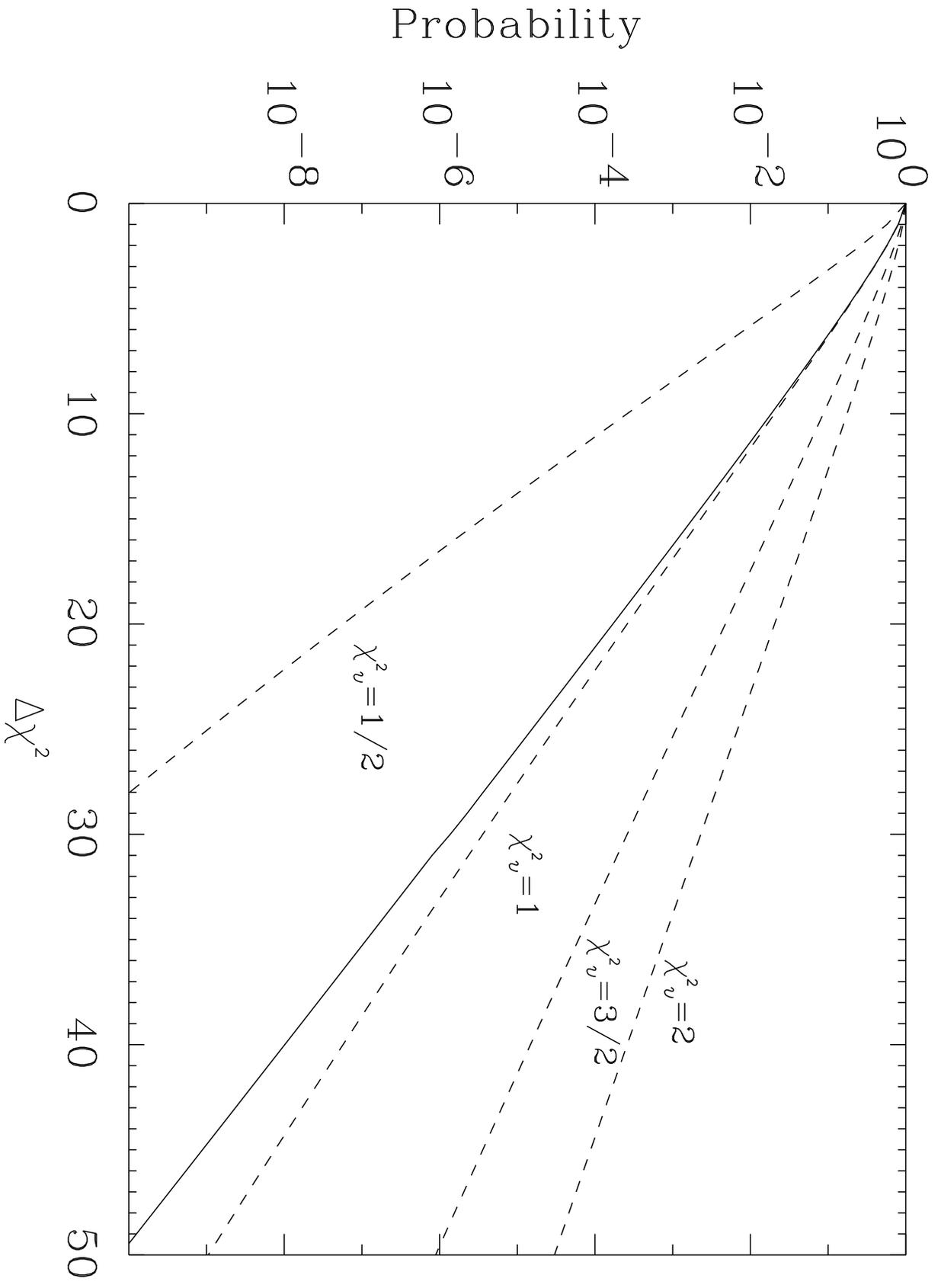
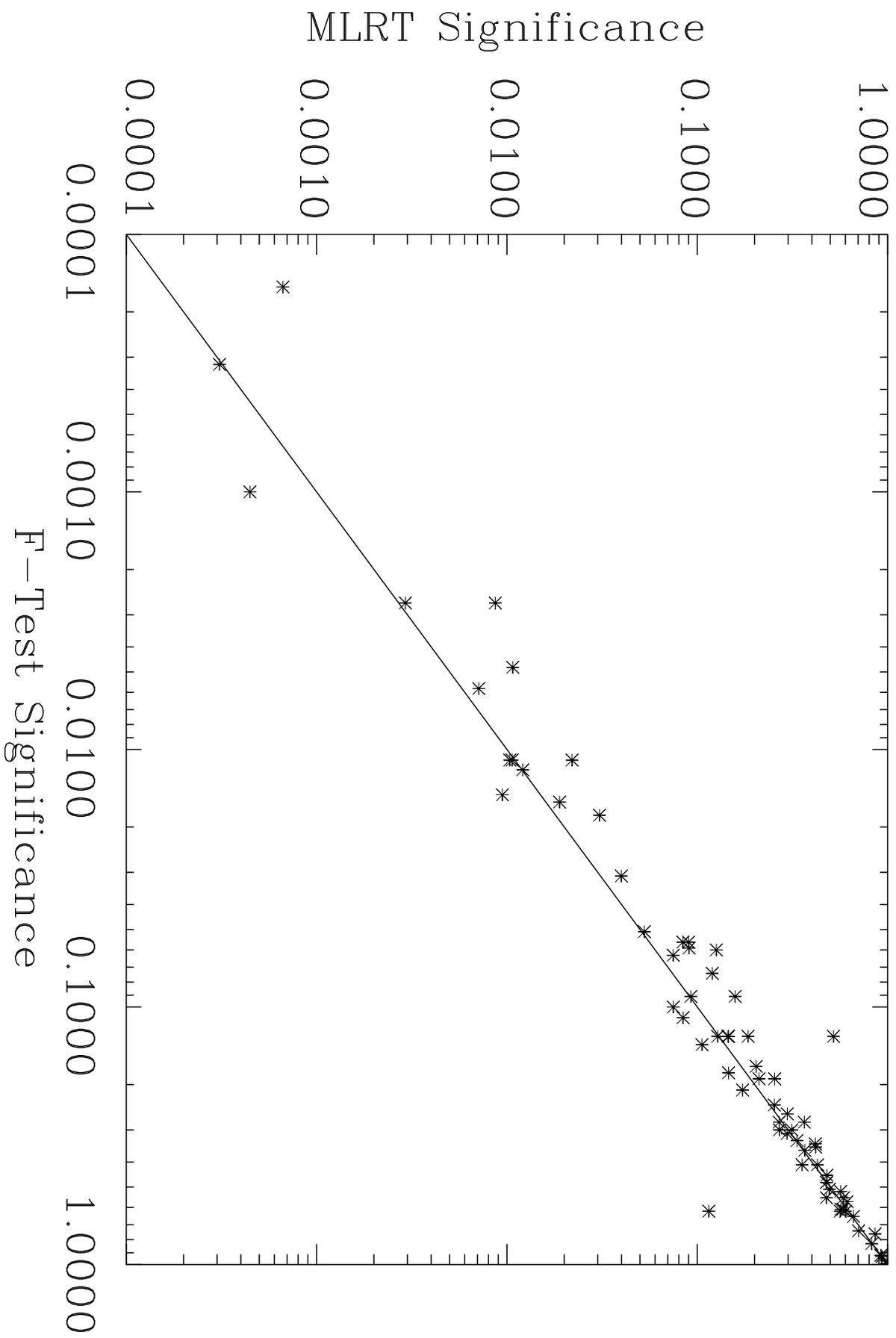
\end{document}